\documentclass[usenatbib]{mnras}

\usepackage{graphicx}
\usepackage{amssymb}
\usepackage{comment}
\usepackage{wallpaper}
\usepackage[normalem]{ulem}

\begin{document}

\title[Anisotropy in Galaxy Encounters]{Effects of Halo Anisotropy on Disc Galaxy Encounters}

\author[T. Umiamaka \& J. E. Barnes]{Tatum M.~Umiamaka$^{1,2}$, Joshua E.~Barnes$^3$
\\
$^1$Department of Physics and Astronomy, University of Hawai`i at M{\=a}noa; \textsf{tatumumi@hawaii.edu}\\
$^2$Department of Astronomy, California Institute of Technology; \textsf{tumiamak@caltech.edu}\\
$^3$Institute for Astronomy, University of Hawai`i at M{\=a}noa; \textsf{barnes@hawaii.edu}}

\date{\today}

\label{firstpage}
\pagerange{\pageref{firstpage}--\pageref{lastpage}}

\maketitle

\begin{abstract}
\noindent
Spherical galaxy models with radially anisotropic velocity distributions merge faster than their isotropic counterparts.  Here we investigate the effects of radially anisotropic haloes on the dynamics of disc galaxy encounters. We use stable galaxy models with isotropic bulges, thin rotating discs, and dark haloes which are either isotropic or radially anisotropic.  Our simulations confirm that anisotropy can markedly accelerate orbit decay in galaxy interactions; in particular, radially anisotropic haloes transfer a good deal of orbital angular momentum to internal motions even \textit{before} the galaxies reach their first pericentre. Consequently, for a given initial orbit, the anisotropic models undergo closer and more violent interactions, and their discs generate more massive tidal features. If real disc galaxies have radially anisotropic haloes, our findings may have implications for estimated merger time-scales and remnant morphology.
\end{abstract}


\section{Introduction}
\label{sec:introduction}

\noindent
The dynamics and evolution of interacting galaxies have been studied intensively since the 1970's. \cite{TT1972} did much to launch the subject  with their paper on the origin of galactic bridges and tails. They used highly idealized models of interacting galaxies -- each consisting of a massive central particle surrounded by a disc of test particles -- to argue that the bridges and tails in binary galaxies are tidal relics of close encounters. Since two massive particles can only follow Keplerian orbits, \citeauthor{TT1972} could not reproduce the process of orbit decay. \cite{White1979} remedied this by employing a few hundred massive particles per galaxy; in his simulations, the interacting galaxies could tidally deform each other, transferring orbital angular momentum to inertial degrees of freedom. Since then, most parameters that influence the dynamics and evolution of galaxy encounters have been well studied. However, the possibility that interacting galaxies have anisotropic haloes has received relatively little attention.

Most self-consistent simulations of disc galaxy encounters have used isotropic or nearly isotropic haloes \citep[e.g.,][]{NW1983, Barnes1988, Barnes1992, Hernquist1992, Hernquist1993, DMH1996, SW1999, NB2003, Barnes2016}. These simulations produced bridges and tails similar to those seen the \cite{TT1972} calculations. They also showed that extended dark haloes encourage orbit decay: models of bright ($\sim L_{*}$) galaxies typically merge within $\sim 1$~Gyr of their initial encounter. This timescale has implications for the overall merger rate inferred from observations, and for interpretation of stellar populations formed by merger-induced starbursts. In addition, the remnants produced by these simulated mergers share many features with observed merger remnants. One interesting discrepancy is that simulated remnants produced by galaxies with massive, compact haloes rapidly reaccrete their tidal tails \citep{DMH1996, SW1999} whereas objects such as NGC~7252 retain their tails well after the galaxies merge \citep{Schweizer1982}.

The preference for isotropic haloes in the above studies was largely motivated by simplicity. This choice may be too restrictive, given the paucity of observational information on halo velocity distributions. Haloes formed in $\Lambda$CDM simulations typically exhibit radial anisotropy due to gravitational collapse \citep{WLGM2015}. However, only a few authors have incorporated haloes with radially or tangentially anisotropic velocity distributions into studies of galaxy encounters. \cite{MAD2007} studied mergers of disc galaxies with mildly anisotropic haloes and found no correlation between the remnant anisotropy and initial halo anisotropy. \cite{AB2007} investigated dynamical friction in elliptical galaxies and observed that satellites orbiting in radially anisotropic host galaxies have slightly shorter fall times. \cite{VBE2022} found that the radial anisotropy of the host galaxy inhibits the circularization of the satellite galaxy's orbit. Additionally, \cite{RFSMPI2022} and \cite{Vasiliev2024} both included anisotropic haloes in studies of the Milky Way and the Large Magellanic Cloud (LMC) system. They find radially anisotropic haloes are more sensitive to the LMC's gravitational field than their isotropric counterparts. \cite{Vasiliev2024} showed the LMC's present position and velocity are consistent with an initial orbit with greater angular momentum if the dark halo is radially anisotropic.

These studies highlight the value of an in-depth investigation into mergers of galaxies containing anisotropic dark haloes. \cite{SB2023} explore the impact of radial anisotropy on orbit decay between spherical galaxies. Their findings suggest that anisotropic haloes are more susceptible to tidal deformation, leading to faster orbit decay. This is a potentially strong effect, reducing the decay time-scale by up to 50\% for highly anisotropic systems. Radially anisotropic haloes deform each other even during the initial approach, so the choice of starting time has a measurable influence on the outcome of encounters of anisotropic systems.

The goal of our investigation is to see if the effects reported by \cite{SB2023} also occur with halo-dominated disc galaxy models. We are also curious to see how the anisotropy of the dark haloes influences the morphology and evolution of the stellar bridges and tails. We outline our N-body methodology in \S~\ref{sec:methods}, and present numerical results in \S~\ref{sec:results}. Discussions and conclusions are presented in \S~\ref{sec:discussion} and \S~\ref{sec:conclusion}.  An appendix presents some results on the radial orbit instability in a bulge/disc/halo galaxy model.

\section{Methods}
\label{sec:methods}

 Our approach is based on \cite{Barnes2016}, replacing the isotropic haloes used there with radially anisotropic counterparts. Below we emphasize the steps needed to set up the anisotropic haloes and refer the reader to \cite{Barnes2016} for other details. We describe our galaxy model (\S~\ref{sec:mass_model}, \S~\ref{sec:velocity_distribution}), simulation technique (\S~\ref{sec:simulations}), test model stability (\S~\ref{sec:stability}), set up encounters (\S~\ref{sec:encounters}), extract the trajectories (\S~\ref{sec:trajectories}), and measure tidal fractions (\S~\ref{sec:tidal_material}).

We use an arbitrary system of units in which the gravitational constant is numerically equal to unity. In other words, we set $G = 1 \, \mathcal{M}^{-1} \, \mathcal{L}^3 \, \mathcal{T}^{-2}$, where $\mathcal{M}$, $\mathcal{L}$, and $\mathcal{T}$ represent our units of mass, length, and time, respectively. Choosing physical values for any two of these units automatically determines the third. For example, if we set $\mathcal{M} = 2.5 \times 10^{11} \, M_{\odot}$ and $\mathcal{L} = 30 \, \mathrm{kpc}$ then the corresponding unit of time is $\mathcal{T} = 203.5 \, \mathrm{Myr}$.

\subsection{Mass Model}
\label{sec:mass_model}

\begin{figure}
  \centering
  \includegraphics[width=0.9\linewidth]{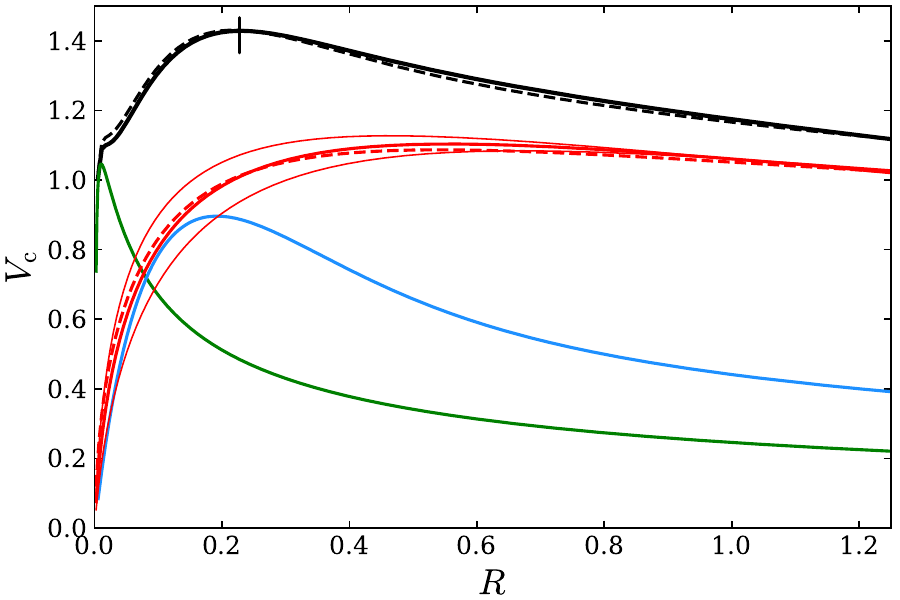}
  \caption{Circular velocity curves for our chosen model and alternate haloes. The bulge, disc, and halo curves are shown in green, blue, and red, respectively, with the total curves in black. NFW model curves are dashed; alternate Einasto halo models are represented with thinner curves.} 
  \label{fig:fig01}
\end{figure}

The model has three collisionless components: a dark halo, a disc, and a central bulge. The halo is $90$~percent of the galaxy's mass, the disc accounts for $7.5$~percent, and the  bulge for $2.5$~percent. Fig.~\ref{fig:fig01} shows circular velocity curves for our chosen model. The total rotation curve  is displayed in black, while contributions from the bulge, disc, and halo are shown in green, blue and red, respectively.

\textbf{1.} We initially planned to use a NFW \citep{Navarro1997} model for our dark halo, similar to \cite{MAD2007} and \cite{Barnes2016}.  However, an NFW profile must be tapered at large radii to render its total mass finite. This tapering severely limits the amount of anisotropy that can be imposed without making the distribution function  negative. Since we wanted to use highly anisotropic halo models, we opted instead to use an \cite{Einasto1965} model, which has a finite mass and does not require tapering.

The \cite{Einasto1965} density profile is
\begin{equation}
    \rho_\mathrm{h}(r) = \frac{d^{3n}}{4 \pi n \Gamma (3n)} \frac{M_\mathrm{h}}{r^{3}_\mathrm{h}} \exp \left[-d\left( \frac{r}{r_\mathrm{h}}\right)^{1/n} \right]
\end{equation}
where the dimensionless quantity $d$ satisfies $\Gamma (3n,d) = \frac{1}{2} \Gamma (3n).$  This profile has three parameters: the total mass $M_\mathrm{h} = 2.25$, the half-mass radius $r_\mathrm{h} = 1$, and the Einasto index $n$. The first two fix the scale of the halo, while $n$ controls the shape of the density profile.

We intend our Einasto halo to be a drop-in replacement for the isotropic NFW halo. To this end, we need to find an Einasto index $n$ which closely mimics the NFW profile. In Fig.~\ref{fig:fig01} we display the circular velocity curves for three Einasto haloes (solid red curves) with indices $n = 3, 3.5, 4$, as well as an NFW halo (dashed red curve). The Einasto model with index $n = 3.5$ (represented by the heaviest curve) provides a good approximation to the rotation curve of our original NFW model; we use this choice below.

\textbf{2.} Our disc has an exponential radial profile \citep{deVaucouleurs1959, Freeman1970} and an isothermal vertical profile \citep{Spitzer1942, VS1981},
\begin{equation}
    \rho_\mathrm{d}(R,z) = \frac{M_\mathrm{d}}{4\pi \alpha^2_\mathrm{d} z_\mathrm{d}}e^{-R\alpha_\mathrm{d}} \mathrm{sech}^2(z/z_\mathrm{d}). 
\end{equation}
The disc mass is $M_\mathrm{d} = 0.1875$, the inverse scale length is $\alpha_\mathrm{d} = 12.0$ and the scale height is $z_\mathrm{d}$. We fix the scale height at $z_\mathrm{d} = 0.125/\alpha_\mathrm{d}$ for all models; consequently, our discs are quite thin.

\textbf{3.} For the bulge, we adopt a \cite{Jaffe1983} profile
\begin{equation}
    \rho_\mathrm{b}(r) = \frac{M_\mathrm{b} a_\mathrm{b}}{4 \pi r^2 (a_\mathrm{b} + r)^2}.
\end{equation}
The bulge mass is $M_\mathrm{b} = 0.0625$, and we fix the scale radius at $a_\mathrm{b}=0.04$ for all our simulations. This bulge is compact, which helps suppress bar instabilities in the disc. For numerical convenience, we exponentially taper the asymptotic $\rho \propto r^{-4}$ density profile at $100$ times the scale radius as in \cite{Barnes2016}.

With these parameter choices, our galaxy model broadly resembles the Milky Way.  It has a somewhat more compact halo, yielding the gradually falling total rotation curve (black) shown in Fig.~\ref{fig:fig01}. At a radius of three scale lengths, roughly corresponding to the Sun's position in the Milky Way, the circular velocity is $v_c = 1.427$ and the corresponding orbital period is $T_{\mathrm{orb}} = 1.10$. Adopting the scaling to physical units discussed above, this orbit period is $224 \, \mathrm{Myr}$.

\subsection{Velocity Distributions}
\label{sec:velocity_distribution}

As in \cite{BH2009} we initialize the bulge and halo as spherical systems in mutual equilibrium with a spherical approximation to the potential of the disc. The potential $\Phi(r)$ is generated by the combined halo, bulge, and sphericized disc. We smooth the total density profile to account for gravitational softening \citep{Barnes2012}, and numerically compute $\Phi(r)$ using Poisson's equation.

\textbf{1.} We use an Osipokov-Meritt distribution function (DF) for the dark halo \citep{Osipkov1979, Meritt1985}. This DF has the form $f_\mathrm{h}(\mathbf{r},\mathbf{v}) = f_\mathrm{h}(Q)$, in which $Q = E + \frac{1}{2} J^2 / r_a^2$, where $E\equiv \frac{1}{2}|v|^2+\Phi(|r|)$ is the specific binding energy, $J$ is the specific angular momentum, and $r_{\mathrm{a}}$ is the anisotropy radius, which controls the degree of anisotropy. If we resolve the velocity into its radial component $v_r$ and tangential component $v_t$, we obtain
\begin{equation}
    Q = \frac{1}{2}v_r^2 + \frac{1}{2}\left( 1+\frac{r^2}{r_a^2}\right)v_t^2 + \Phi(|r|)
\end{equation}
If $r \ll r_a$ then $Q \simeq E$ and the velocity distribution is approximately isotropic. If $r \gg r_a$ the transverse velocities are suppressed more than radial velocities, and the distribution exhibits a pronounced radial preference. We quantify the degree of anisotropy at radius r using
\begin{equation}
    \beta (r) \equiv 1- \frac{\langle v_t^2 \rangle}{\langle 2v_r^2 \rangle} = \frac{r^2}{r^2+r_a^2}
\end{equation}
(the last equality assumes an Osipkov-Merritt DF). If $\beta = 0$ the distribution is isotropic, while if $0 < \beta \leq 1$ the distribution is radially anisotropic. 
The DF is computed from the density $\rho (r)$, potential $\Phi (r)$, and anisotropy radius $r_{\mathrm{a}}$:
\begin{equation}
    f_\mathrm{h}(Q) = \frac{1}{\sqrt{8} \pi ^2} \frac{d}{dQ} \int_{Q}^{0} \frac{1}{\sqrt{\Phi - Q}} \frac{d\tilde{\rho}_\mathrm{h}}{d \Phi} d \Phi,
\end{equation}
 where $\tilde{\rho}_\mathrm{h}$ is given by
 \begin{equation}
    \tilde{\rho}_\mathrm{h} = \left( 1+\frac{r^2}{r_a^2}\right)\rho_\mathrm{h} (r).
\end{equation}
The potential $\Phi (r)$ can not be solved analytically, so we employ numerical methods.

\textbf{2.} The disc is initialized using an approximate DF \citep{BH2009}.  Motion in the vertical direction is assumed to decouple from motion in the disc plane; the former is described by the isothermal sheet model \citep{Spitzer1942}, while the latter is constrained by the Jeans equations, including a treatment of asymmetric drift \citep{BT2008}.  We set the ratio of radial to vertical velocity dispersion $\sigma_R / \sigma_z = 2$, matching the value found in the solar neighborhood.

\textbf{3.} We adopt an isotropic DF for the bulge, using essentially the same formalism used to construct the halo, with $r_{\mathrm{a}} = \infty$.

\subsection{N-Body Simulations}
\label{sec:simulations}

To follow the time evolution of our models, we employ self-consistent N-body simulations. We represent the DF with an assembly of N particles $i = 1, \dots, N$, each with position $\mathbf{r}_i$ and velocity $\mathbf{v}_i$. The mass $m_i$ of a particle depends on which part of the galaxy  it represents. Halo particles have masses of $m_\mathrm{h} = 2^{-16}$, while disc and bulge particles have masses of $m_\mathrm{d} = m_\mathrm{b} = 2^{-18}$. A single galaxy therefore has $N_\mathrm{h} = M_\mathrm{h} / m_\mathrm{h} = 147456$ halo particles, $N_\mathrm{d} = M_\mathrm{d} / m_\mathrm{d} = 49152$ disc particles, and $N_\mathrm{b} = M_\mathrm{b} / m_\mathrm{b} = 16384$ bulge particles.

We use Monte-Carlo sampling to choose the initial coordinates of each particle. The probability that particle $i$ has initial position $\mathbf{r}_i$ and initial velocity $\mathbf{v}_i$ is proportional to the value of the DF at those coordinates.  This introduces some uncertainty, since the number of particles in a phase-space volume $V$ will have fluctuations of order $\sqrt{N_V}$ around its mean value of $N_V$. These fluctuations can only be reduced by increasing the total number of particles simulated, which is computationally expensive. We therefore run multiple realizations of each simulation, and interpret the variations between realizations as a measure of the intrinsic uncertainty of our experiments.

Gravitational interactions between particles are calculated using a tree algorithm \citep{BH1986}.  We adopted an opening angle of $\theta = 0.8$ and included quadrapole terms in calculating the gravitational field of each cell \citep{Hernquist1987}.  This yields accelerations with RMS relative errors of $\sim 0.06$~per cent.

Particle trajectories are computed using a time-centered leapfrog method.  The time-step is $\delta t = 1/1024$.  We use Plummer softening \citep[e.g.,][]{Barnes2012} to suppress short-range interactions. The softening length, $\varepsilon = 0.0025$, is smaller than the vertical scale height of the disc and the scale radius of the bulge, although the inner $r^{-2}$ density profile of the latter is only marginally resolved.  This represents a compromise between spatial resolution and simulation cost; a smaller $\varepsilon$ would improve the resolution, but a shorter time-step would be needed to ensure good energy conservation.  The present level of resolution appears adequate, as our main interest is the large-scale response of the models to external tidal fields.  All simulations were run on 4 GHz Intel processors with single precision arithmetic; a typical experiment spanning $16$ time units took $\sim 75$ hours of processor time.

\subsection{Stability}
\label{sec:stability}

Before conducting merger simulations, we checked the stability of our galaxy models.   Discs are notoriously prone to bar instabilities, although this is unlikely to be a concern here since our model is based on a bar-stable model \citep{Barnes2016}.  However, the anisotropic haloes used in our simulations may be susceptible to the radial orbit instability \citep{Antonov1973, MA1985, BGH1986}, which spontaneously transforms an initially spherical equilibrium system into an elongated triaxial configuration.  While avoiding the radial orbit instability, we wish to make our haloes highly anisotropic, so as to unambiguously demonstrate the effects of halo anisotropy.  The degree of halo anisotropy increases as the anisotropy radius $r_\mathrm{a}$ is reduced; we therefore want to find a value of $r_\mathrm{a}$ which is just slightly larger than values which produce instability.

In practice, this is very similar to \cite{SB2023}, and we follow their general approach.   Preliminary tests with pure \cite{Einasto1965} models placed the stability boundary near $r_\mathrm{a} = 0.5$, so we chose $r_{\mathrm{a}}$ values of $2^{-3/2}$, $2^{-1}$, $2^{-3/4}$, and $2^{-1/2}$.  These all yield halo DFs which are everywhere positive, so the resulting models are physically consistent.  We ran four independent realizations for each value of $r_\mathrm{a}$ to more accurately distinguish between stable and unstable models.  To quantify halo shapes, \citeauthor{SB2023} evaluated the moment of inertia tensor for halo particles between the $75^\mathrm{th}$ and $87.5^\mathrm{th}$ percentiles in initial binding energy; they computed eigenvalues $\lambda_1 > \lambda_2 > \lambda_3$ of this tensor, and interpreted $\sqrt{\lambda_3 / \lambda_1}$ as the axial ratio $(c/a)$ for this sample of halo particles.  We do likewise, and also examine the angle $\theta_\mathrm{minor}$ between the halo minor axis (eigenvector $\mathbf{e}_3$) and the disc's axis of symmetry, since the latter's non-spherical gravitational field may influence the halo's orientation.

\begin{figure} 
  \centering
  \includegraphics[width=1\linewidth]{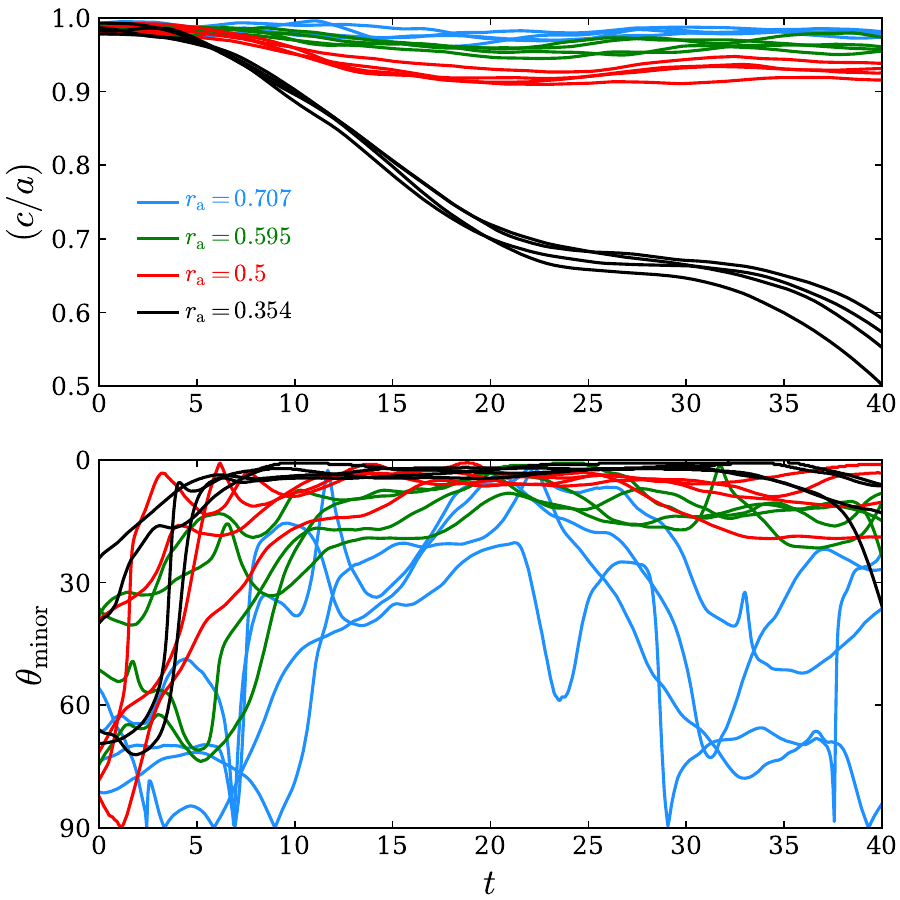}  
  \caption{Halo stability of simulations with varying $r_\mathrm{a}$ values. The top panel shows the halo axial ratio $(c/a)$ over time, while the bottom panel depicts the evolution of the angle $\theta_\mathrm{minor}$ between the disc symmetry axis and the halo's minor axis. Four realizations are shown for each anisotropy radius $r_{\mathrm{a}}$.} 
  \label{fig:fig02} 
\end{figure}

Fig.~\ref{fig:fig02} presents the results of the halo stability tests. As functions of time, the top panel shows the halo axial ratio $(c/a)$, while the bottom panel shows the angle $\theta_\mathrm{minor}$.  The four experiments with $r_\mathrm{a} = 0.354$ (black curves) are clearly unstable; in every case, the haloes rapidly lose their initial spherical shapes, flattening by factors of up to $2$ along the disc's symmetry axis. In contrast, the runs with $r_\mathrm{a} = 0.707$ (blue) appear stable, maintaining halo axial ratios $(c/a) \gtrsim 0.95$; the small distortions we observe appear randomly oriented, showing little or no influence from the disc's gravitational field.

Between these extremes are the simulations with $r_\mathrm{a} = 0.595$ (green) and $r_\mathrm{a} = 0.5$ (red).  Neither is as close to spherical as the $r_\mathrm{a} = 0.707$ sample, with the more anisotropic haloes exhibiting larger distortions.  Yet neither displays the rapid evolution of the unstable haloes; they appear to settle into slightly flattened configurations with minor axies fairly well-aligned with the symmetry axis.   Stable systems with high levels of radial anisotropy respond more strongly to non-spherical perturbations than do their more isotropic counterparts \citep[e.g.,][]{RFSMPI2022, Vasiliev2024, SB2023}; it's likely that the mild flattening of these haloes arises in response to the disc's non-spherical field.  Given that this flattening remains constant after an initial period of adjustment, we infer that these models are stable, and adopt $r_\mathrm{a} = 0.5$ for our anisotropic halo models. Further details can be found in the Appendix.

We visually checked the above simulations to see if the discs develop bars. While we observe transient spiral features, we do not find any long-lived bars.

\subsection{Encounter Sample}
\label{sec:encounters}

Our encounter sample is motivated by the following considerations:
\begin{enumerate}

    \item To examine the effects of anisotropy, each encounter is simulated with both isotropic ($r_\mathrm{a} = \infty$) and anisotropic ($r_\mathrm{a} = 0.5$) haloes.

    \item We set up all the encounters by launching a pair of galaxy models towards each other on initially parabolic orbits.  To examine the evolution of encounters with a modest range of orbital angular momenta, these orbits have pericentric separations of $r_\mathrm{p} = 0.25$, $0.5$, and $1.0$.

    \item At a minimum, an encounter simulation should start early enough to insure that the galaxies initially interact much like two point masses.  \cite{SB2023} found that anisotropic systems can undergo significant orbit decay during their initial approach, and that earlier starting times increase this effect. To study this, we vary the starting time.

    \item The tidal response of the discs depends on the encounter geometry \citep{TT1972}.  We employ four different configurations which between them sample eight different disc orientations from prograde (spin and orbit parallel) to retrograde (antiparallel).

    \item To estimate run-to-run variations, a subset of our experiments are repeated $N_\mathrm{real} = 2$ or $4$ times. A different random-number seed was used to initialize each simulation.

\end{enumerate}

The initial positions and velocities of the galaxies are derived from the ideal parabolic orbits of two point-mass bodies, each with mass $M = 2.5$.  We adopt a coordinate system in which the bodies move clockwise in the $x$-$y$ plane, and define the instant of pericentre as time $t = 0$. At that time, the two bodies are located at $\pm r_\mathrm{p} / 2$ along the $y$-axis.

We choose starting times $t_\mathrm{s} < 0$ to place the galaxies on the inbound leg of their initial orbit. This choice entails a compromise.  Very early starting times better approximate the ideal of two galaxies falling from an infinite distance, consistent with their parabolic initial orbit; this also allows more time for any orbit evolution which may occur during the initial approach.  However, starting early has drawbacks; in addition to the obvious consideration of computing time, run-to-run variations in the initial trajectories become larger \citep{SB2023}, and the discs have more time to heat up before interacting with each other.

Our default choice of $t_\mathrm{s} = -6$ assigns the galaxies an initial separation of $r_\mathrm{s} = 8.85$.  The degree of overlap at this initial distance can be inferred by comparing the gravitational potential energy $U_\mathrm{tot}$ of the initial system to the value it would have if the galaxies were completely disjoint, $U_1 + U_2 - G M^2 / r_\mathrm{s}$, where $U_j$ is the potential energy of galaxy $j$ in isolation \citep{SB2023}.  For $t_\mathrm{s} = -6$, the difference is $\sim 0.6$~per cent of $G M^2 / r_\mathrm{s}$, implying a modest degree of overlap.  We ran additional simulations with $t_\mathrm{s} = -10$ and $-14$, yielding initial separations $r_\mathrm{s} = 12.62$ and $15.91$, respectively; the corresponding offsets in potential energy amount to $\sim 0.2$ and $\sim 0.05$~per cent of $G M^2 / r_\mathrm{s}$.

The orientation of each disc is specified by two rotation angles.  A disc is created spinning clockwise in the x-y plane; it is first rotated by $\theta_\mathrm{x}$ about the x-axis, and then by $\theta_\mathrm{z}$ about the z-axis.  Thus, $\theta_\mathrm{x}$ sets the disc's inclination with respect to the orbital plane, while $\theta_\mathrm{z}$ determines the argument of pericentre \citep{TT1972}.  Table~\ref{tab:configurations} lists four configurations, based on the symmetry axes of a tetrahedron, which coarsely span the space of possible orientations \citep[e.g.,][]{Barnes1992, MAD2007}.  Configuration~A defines a prograde encounter, with one disc in the orbit plane and the other inclined by $71^\circ$, while configuration~B reverses the spins of both discs.  Configurations~C and~D pair a prograde disc with a retrograde disc.

\begin{table}
  \centering
  \caption{Disc configurations for encounter simulations.}
  \label{tab:configurations}
  \begin{tabular}{ccccc}
  \hline
  & \multicolumn{2}{c}{Disc~1} & \multicolumn{2}{c}{Disc 2} \\
  & $\theta_x$ & $\theta_z$ & $\theta_x$ & $\theta_z$ \\
  \hline
  A & $0^\circ$ & $0^\circ$ & $71^\circ$ & $60^\circ$\\
  B & $180^\circ$ & $0^\circ$ &  $251^\circ$ & $60^\circ$\\
  C & $71^\circ$ & $300^\circ$ & $251^\circ$ & $180^\circ$\\
  D & $71^\circ$ & $180^\circ$ & $251^\circ$ & $300^\circ$ \\
  \end{tabular}
\end{table}

Table~\ref{tab:experiments} summarizes the encounters used in this study.  With two choices for the anisotropy radius $r_\mathrm{a}$, three choices for the pericentric separation $r_\mathrm{p}$, four configurations, and three choices for starting time $t_\mathrm{s}$, we have $72$ possible parameter combinations.  That seemed excessive, especially since we want to run multiple realizations of many encounters.  We opted instead to use the $r_\mathrm{p} = 0.5$ encounters to explore the effects of $t_\mathrm{s}$, with configuration~A spanning the greatest range of values.  All told, we ran a total of $68$ simulations, sampling $34$ different parameter combinations.

\begin{table}  
  \centering
  \caption{Encounter parameters.  Each encounter was run with both anisotropic ($r_\mathrm{a} = 0.5$) and isotropic haloes.  We ran encounters with pericentric separation $r_\mathrm{a} = 0.5$ using multiple values of the starting time $r_\mathrm{s}$ as shown, with the corresponding number of $N_\mathrm{real}$ realizations.}
  \label{tab:experiments}  
  \begin{tabular}{cccccccc}
  \hline
     $r_{\mathrm{p}}$ & Config & \multicolumn{3}{l}{$t_\mathrm{s}$}  & \multicolumn{3}{l}{$N_\mathrm{real}$}  \\
    \hline
     0.25 & A & -6 &    &      & 2 &   &  \\
          & B & -6 &    &      & 1 &   &  \\
          & C & -6 &    &      & 1 &   &  \\
          & D & -6 &    &      & 1 &   &  \\
     0.5  & A & -6 & -10 & -14 & 4 & 4 & 4\\
          & B & -6 & -10 &     & 2 & 2 &  \\
          & C & -6 & -10 &     & 2 & 2 &  \\
          & D & -6 & -10 &     & 2 & 2 &  \\
     1.0  & A & -6 &    &      & 2 &   &  \\
          & B & -6 &    &      & 1 &   &  \\
          & C & -6 &    &      & 1 &   &  \\
          & D & -6 &    &      & 1 &   &  \\
          \hline
  \end{tabular}
\end{table}

\subsection{Trajectories}
\label{sec:trajectories}

\begin{figure}
    \centering
    \includegraphics[width=\linewidth]{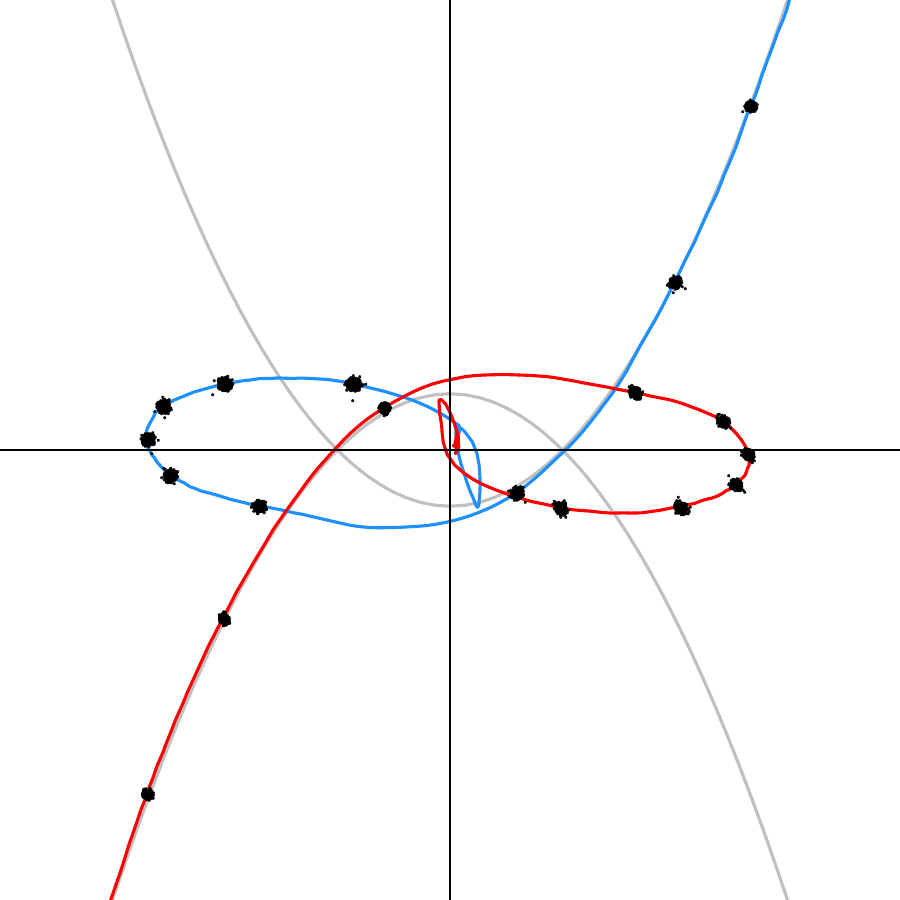}\\[0.25in]
    \includegraphics[width=\linewidth]{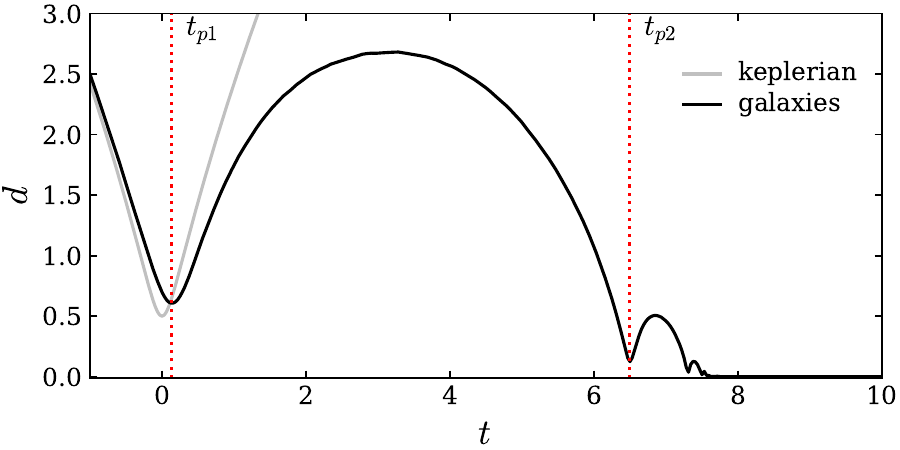}
    \caption{Trajectories of merging galaxies (top) and the corresponding radial trajectory (bottom). Above, the blue and red curves show actual trajectories, black points show positions of bulge particles used to determine these trajectories, and gray curves show the initial parabolic orbit. Below, the solid black line shows the actual radial trajectory, the dashed black curve represents the initial parabolic trajectory, and the dotted red lines indicate the first $t_{\mathrm{p1}}$ and second $t_{\mathrm{p2}}$ pericentres. Simulation parameters: anisotropy $r_\mathrm{a} = \infty$, pericentre $r_\mathrm{p} = 0.5$, starting time $t_\mathrm{s} = -6$, disc configuration A.}
    \label{fig:fig03}
\end{figure}

At each time-step, we measure the position of each galaxy from the centroid of a fixed sample of tightly-bound bulge particles.  Specifically, we order bulge particles by initial binding energy (computed with respect to their parent galaxy), and use the center-of-mass of the first $N_\mathrm{b}/4 = 4096$ particles.  This scheme is quite robust, as the \citeauthor{Jaffe1983}-model bulges employed in these simulations are very compact, and their most tightly-bound quartiles are barely perturbed prior to actual merger.  Galaxy positions determined in this fashion are consistent with the positions inferred from visual inspections of the simulations.  Comparing sample mean and median coordinates, we estimate that our trajectories are accurate to $\pm 10^{-4}$ length units during the encounter phase, and only a few times less accurate post-merger.

The top panel of Fig.~\ref{fig:fig03} presents  a typical encounter trajectory. The bulge particles used to determine each galaxy's position are plotted at unit time intervals; smooth curves show trajectories at intervening times. The galaxies first approach on approximately parabolic orbits, pass each other at first pericentre, come to a near standstill at apocenter, and fall back for a much closer second pericentre. The interaction between the galaxies is encapsulated by the plot in the bottom panel, where we display the distance between the galaxies as a function of time. Local minima occur at pericentres; we will use the time $\Delta t = t_\mathrm{p2} - t_\mathrm{t1}$ between first and second pericentres as an estimate of the orbit-decay timescale. We consider the galaxies to have merged after their third pericentre. By this stage, the discs are deeply intertwined and can no longer be distinguished; the centers of the bulges may still be distinct, but the system as a whole is better described as a single galaxy with a double nucleus.

\subsection{Tidal Material}
\label{sec:tidal_material}

\begin{figure}
  \centering
  \includegraphics[width=\linewidth]{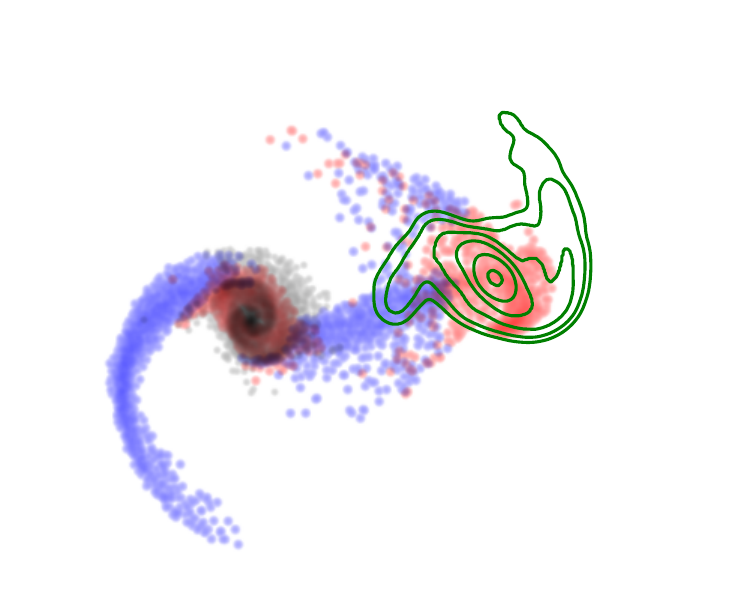}  
  \caption{Classification of tidal features from the in-plane disc in configuration~A.  Particles currently counted as tidal are shown in blue, while those which have either fallen back into their parent galaxy (left) or accreted by the companion galaxy (right) are shown in red.  Particles which have never ventured outside their critical radius $r_\mathrm{tid}$ are shown in black.  Green contours show the companion galaxy's disc.  Simulation parameters: anisotropy $r_\mathrm{a} = \infty$, pericentre $r_\mathrm{p} = 0.5$, starting time $t_\mathrm{s} = -6$, current time $t = 1.5$.} 
  \label{fig:fig04}  
\end{figure}

\cite{Barnes2016} counted a particle as part of a tidal feature if its distance from its parent galaxy exceeded five disc scale lengths, or $r_\mathrm{tid} = 5/\alpha_\mathrm{d} \simeq 0.42$ length units. This works for the strong tidal responses resulting from prograde (e.g., configuration~A) encounters, but it appears to under-estimate the amount of tidal material in retrograde passages. We therefore adopted a different critical radius for each particle: $r_\mathrm{tid} = 0.2 + \hat{r}$, where $\hat{r}$ is the particle's initial orbital radius. A particle is counted as part of a tidal feature the \textit{first} time its current distance from its parent exceeds its critical radius, $r(t) > r_\mathrm{tid}$.  This test is limited to times before second pericentre to count only `classic' tidal bridges and tails raised by the first passage.

Particles which have exceeded their tidal radii may later re-accrete onto either their parent galaxy or its companion. Reaccreted particles, although bound to one galaxy or the other, follow elongated orbits and create diffuse structures, in contrast to bridges or tails. We count a tidal particle as having fallen back the first time its distance to its parent is less than its critical radius, $r(t) < r_\mathrm{tid}$; likewise, a particle transfers if it falls within its critical radius of the companion galaxy, $r'(t) < r_\mathrm{tid}$. In Fig.~\ref{fig:fig04}, we show how this scheme classifies particles from the in-plane galaxy as either tidal (blue) or recaptured (red). The red particles fall into two groups, one surrounding the disc they came from, and the other currently forming a `hook-like' structure slightly below and to the right of the companion. Further evolution of these features is illustrated in section \ref{subsec:results_tidalmaterial}. After second pericentre, this scheme becomes less reliable as some returning particles have pericentric distances exceeding $r_\mathrm{tid}$. We will focus on the evolution of tidal material in bridges and tails before second pericentre to avoid this complication.

\section{Results}
\label{sec:results}

\begin{figure*} 
  \centering
  \includegraphics[width=0.95\textwidth]{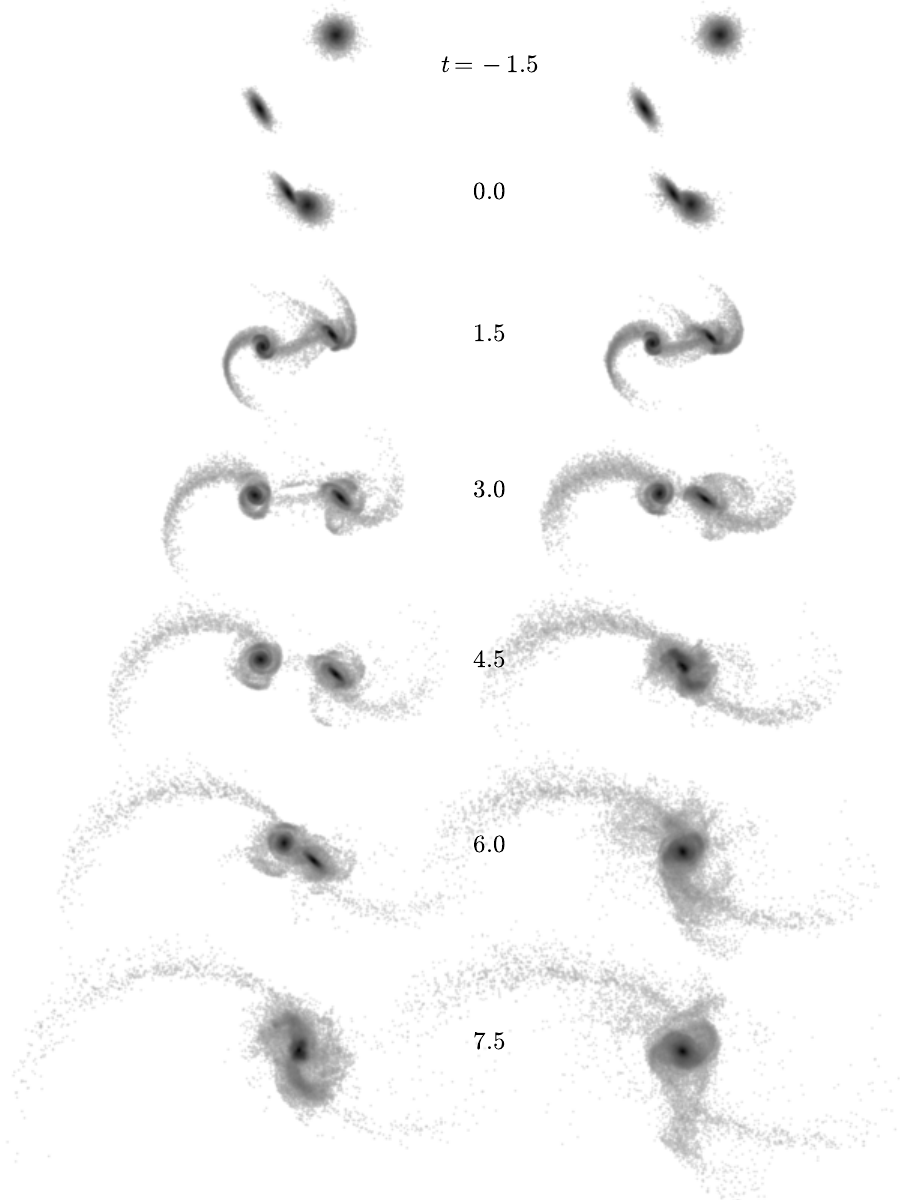}  
  \caption{Evolution of encounters between isotropic (left) and anisotropic (right) galaxies, with pericentric separation $r_\mathrm{p} = 0.5$, starting time $t_\mathrm{s} = -6$, and disc configuration A.}
  \label{fig:fig05} 
\end{figure*}

Does halo anisotropy influence the orbit decay of disc galaxies in the same way that \cite{SB2023} observed for spherical galaxies? How does anisotropy impact the formation and evolution of tidal features? We investigate these questions using a series of controlled simulations in which only the anisotropy of the halo is varied, while all other parameters are held constant. This method ensures direct and meaningful comparisons between isotropic and anisotropic simulations.

Fig.~\ref{fig:fig05} previews our results. Here we compare the evolution of similar encounters between isotropic galaxies (left) and anisotropic galaxies (right) . We display seven pairs of snapshots, with the first taken $1.5$~time units before the initial pericentre and the last captured $7.5$~time units after. The overall scenario in Fig.~\ref{fig:fig05} is repeated in all our experiments; in every case, galaxies reach pericentre, develop tidal features (depending on disc orientation), reach apocenter, fall back for a second and much closer encounter, and merge soon thereafter.

Anisotropy manifests itself in two distinct ways, both visible in Fig.~\ref{fig:fig05}: first, it accelerates orbit decay, and second, it increases the amount tidal material. Initially we see little difference between the two encounters, but after $t = 1.5$, the difference in orbit decay becomes more obvious. Specifically, the anisotropic galaxies (right side) fall together faster than the isotropic ones. For example, the  anisotropic galaxies almost completely merge by $t=4.5$, while the isotropic pair take until $t=7.5$ to coalesce. This correlation between halo anisotropy and orbit decay extends the observations of \cite{SB2023} with spherical galaxies.

\begin{figure*}
  \centering
  \includegraphics[width=1.0\linewidth]{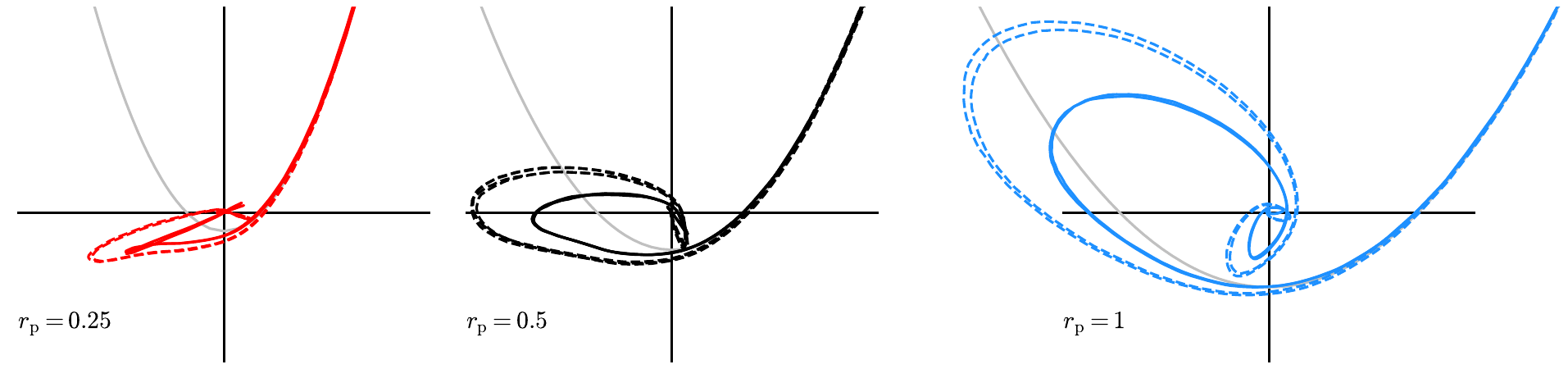}  
  \caption{Relative trajectories of isotropic (dotted) and anisotropic (solid) galaxies in encounters with different $r_\mathrm{p}$ values, as indicated.  All encounters start at $t_\mathrm{s} = -6$ and use  configuration A. For reference, idealized parabolic orbits (gray curves) are shown alongwith the relative trajectories. We include four encounters for each choice of isotropy when $r_\mathrm{p} = 0.5$, and two encounters for each choice at the other $r_\mathrm{p}$ values; some of these curves are hard to distinguish. The horizontal axes are 5 length units from end to end.}
  \label{fig:fig06}  
\end{figure*}

\begin{figure*}
  \centering
  \includegraphics[width=1.0\linewidth]{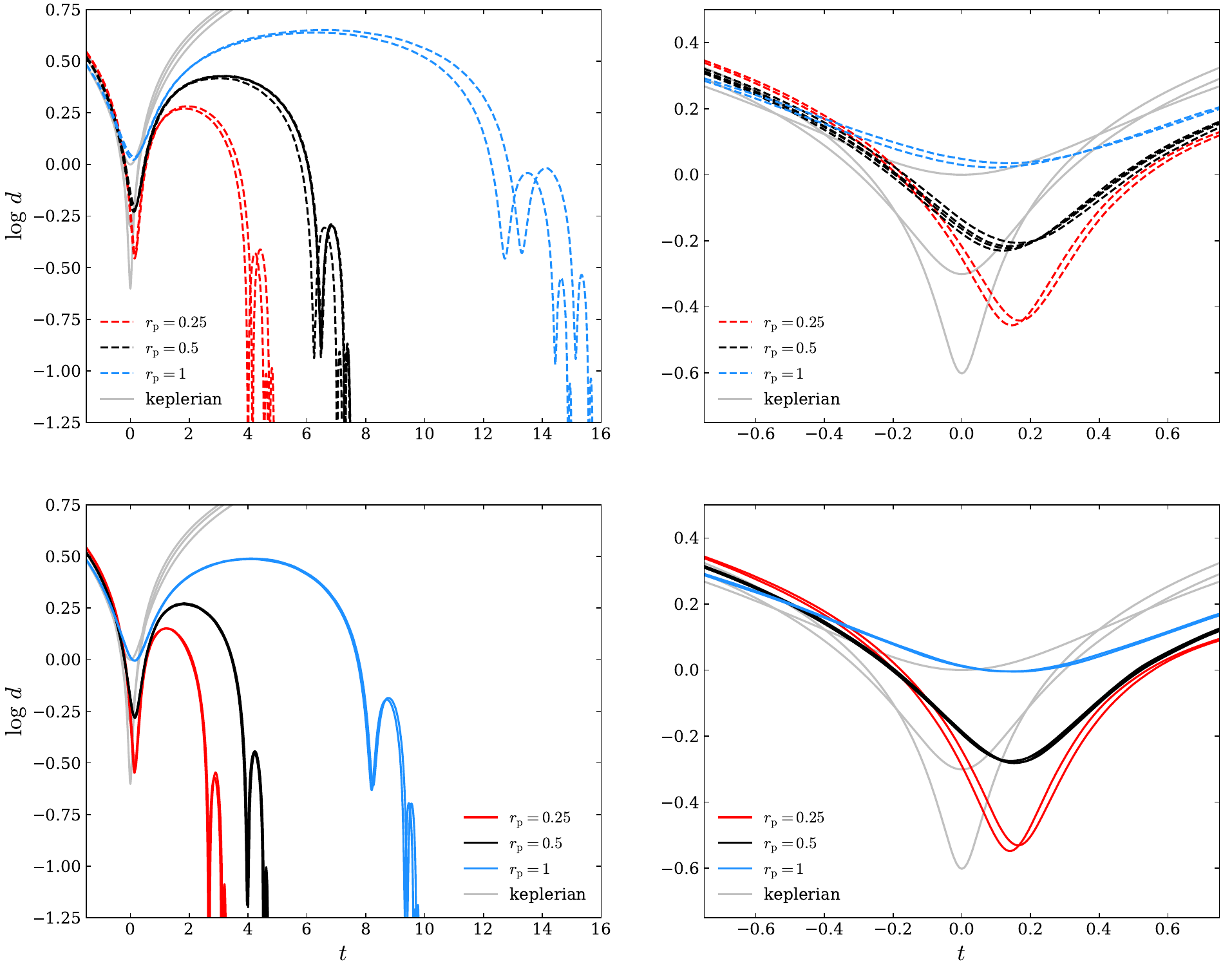}  
  \caption{Radial trajectories of encounters from Fig.~\ref{fig:fig06}.  As in that figure, solid and dotted curves represent encounters of anisotropic and isotropic systems, respectively, while light gray curves show the initial Keplerian parabolas.  Color indicates $r_\mathrm{p}$ value. The right plots show a zoomed-in view of the first pericentre.}
  \label{fig:fig07}  
\end{figure*}

\begin{figure*}
  \centering
  \includegraphics[width=\linewidth]{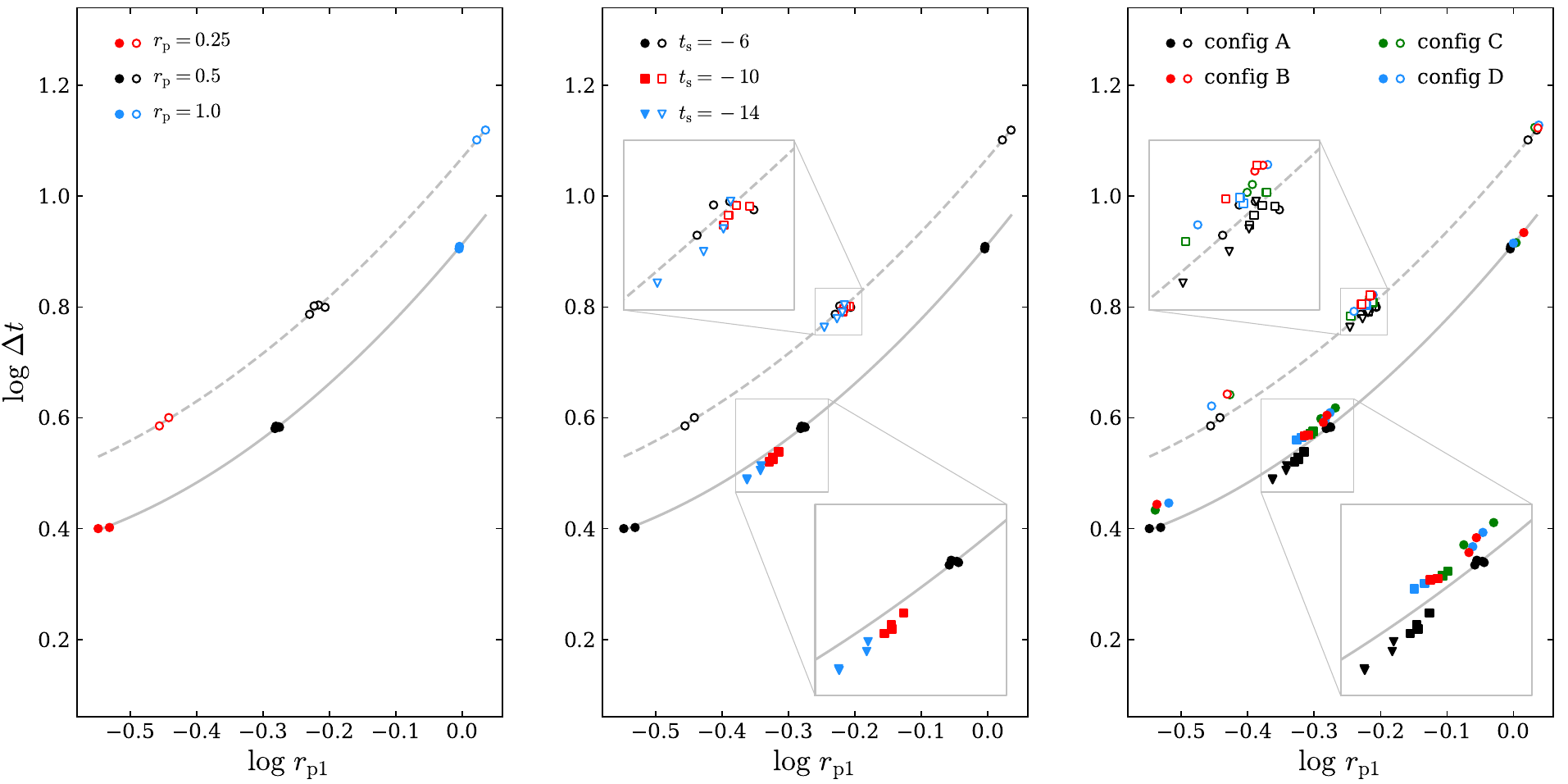} 
  \caption{Relationship between actual pericentric separation $r_\mathrm{p1}$ and orbit decay timescale $\Delta t$.  Solid (open) symbols show results for anisotropic (isotropic) haloes.  Each successive panel builds on the previous one.  Left panel: effect of idealized pericentric separation $r_\mathrm{p}$.  All configuration~A simulations starting at $t_\mathrm{s} = -6$ are shown.  Solid (dashed) curves are second-order fits to the anisotropic (isotropic) results, reproduced identically in all three panels.  Middle panel: effect of starting time $t_\mathrm{s}$.  The insets above and below the main curves zoom in on results for $r_\mathrm{p} = 0.5$.  All configuration~A simulations are shown.  Right panel: effect of disc configuration.  Symbol shapes indicate $t_\mathrm{s}$, as in the middle panel.  All simulations are shown.}
  \label{fig:actualPeri_mergeTime}
\end{figure*}

Differences in tidal features are also more obvious at later times. From $t=1.5$ onward, we observe more tidal material in the encounter between the anisotropic galaxies; in particular, the tails appear better defined and more massive.  However, the overall morphology of the two systems remains similar; the main difference is that the  anisotropic galaxies, as a result of stronger orbit decay, are falling back together by $t = 3$.  This obliterates the bridge between them, and increases the distance from tail-tip to galaxy center.  The anisotropic galaxies have a second pericentre at $t \simeq 4$ and merge soon after, launching fresh plumes of tidal material.  In contrast, the isotropic galaxies don't reach second pericentre until $t \simeq 6.5$; while they gradually fall together, their bridges and tails thin out as tidal material is reaccreted.

\subsection{Orbit Decay}
\label{subsec:results_orbit decay}

Fig.~\ref{fig:fig06} shows how anisotropy and initial orbital angular momentum jointly influence the shapes of orbits.  In general, low-angular momentum orbits (e.g., $r_\mathrm{p} = 0.25$) follow more radial trajectories, while their high-angular momentum counterparts (e.g., $r_\mathrm{p} = 1.0$) are more rounded, avoiding `hairpin' turns.  Anisotropy uniformly makes orbits more radial \citep[cf.][]{VBE2022} and systematically reduces the apocentric separations the galaxies attain between first and second passage.  Qualitatively, an encounter between anisotropic galaxies behaves like an encounter of isotropic galaxies with less angular momentum.  The small loop at the apocenter of the encounter between anisotropic galaxies with $r_\mathrm{p} = 0.25$  is a case in point; \cite{Barnes2016}, using galaxy models very similar to the isotropic models employed here, found such a loop for encounters with $r_\mathrm{p} = 0.125$.  Similar orbital loops were reported by \cite{Tsatsi2015}.  Both studies note that such loops imply the orbital angular momentum temporarily reverses sign.

Moving beyond shapes, Fig.~\ref{fig:fig07} shows how these same orbits evolve in time.  As one might anticipate from the previous figure, anisotropic galaxies, which do not get as far apart after first passage as their isotropic counterparts, also take less time to reach second passage and merge.  For all three choices of pericentric separation $r_\mathrm{p}$, orbit decay times $\Delta t$ are substantially shorter for the anisotropic systems than for the isotropic ones, consistent with \cite{SB2023}.  Closer inspection of the lower panel confirms two additional observations.  First, the self-consistent models reach pericentre slightly later than the equivalent point-masses do, because the gravitational acceleration between two overlapping galaxies is smaller than between two points.  Second, at pericentre the anisotropic models are already somewhat closer together than their isotropic counterparts.  This results because the anisotropic galaxies lose orbital angular momentum even during their initial approach.

Consider a one-parameter family of parabolic encounters, all involving exactly the same galaxy model, which differ only in the choice of initial pericentre $r_\mathrm{p}$.  It's evident that both the actual pericentric separation $r_\mathrm{p1}$ and the orbit decay time $\Delta t$ will be smooth functions of $r_\mathrm{p}$.  Thus, there is an implicit relationship between $\Delta t$ and $r_\mathrm{p1}$.  Closer encounters yield stronger tides, so we also expect $\Delta t$ to increase monotonically with $r_\mathrm{p1}$.

Since we have two galaxy models, our data yield \textit{two} distinct relationships between $r_\mathrm{p1}$ and $\Delta t$: one for the isotropic systems and the other for the anisotropic systems.  The left-hand panel of Fig.~\ref{fig:actualPeri_mergeTime} shows these relationships, with second-order curves fit to the data to guide the eye.  Of particular note is the \textit{vertical} offset between these curves, implying that  anisotropic galaxies with a given actual pericentric separation will reach second pericentre in only $\sim 75$~per cent of the time  isotropic galaxies with the same $r_\mathrm{p1}$ would require. This shows that anisotropy and initial orbit are not entirely degenerate; besides enhancing orbit decay during the initial approach, anisotropy continues to have effects throughout the merging process.

The middle panel of Fig.~\ref{fig:actualPeri_mergeTime} shows how the relationship between the orbit decay time $\Delta t$ and the actual pericentric separation $r_\mathrm{p1}$ is influenced by starting time $t_\mathrm{s}$. Insets above and below the main curves show results for the $r_\mathrm{p} = 0.5$ encounters at a larger scale. Points representing the  encounters of anisotropic galaxies spread out neatly by starting time, whereas their isotropic counterparts display no obvious trend. Starting earlier gives anisotropy more time to have an effect, reducing both $r_\mathrm{p1}$ and $\Delta t$.  Note that the anisotropic points with $t_\mathrm{s} < -6$ fall systematically \textit{below} the curve defined in the left panel, almost as if these simulations were run with a higher level of anisotropy.

The right panel follows the same format as the others but highlights encounters using configurations~B, C, and~D. The influence of disc orientation on orbital evolution is fairly subtle, as previous studies have noted \citep[e.g.,][]{Barnes1992}.  Orbit decay is faster in simulations using configuration~A, which pairs two prograde discs, presumably because the discs undergo more violent tidal interactions.  Consistent with that explanation, this effect is strongest for the closest passages, and apparently independent of halo anisotropy.   

\subsection{Evolution of Tidal Features}
\label{subsec:results_tidalmaterial}

As Fig.~\ref{fig:fig05} already shows, anisotropy influences tidal features. Why is this true? Does anisotropy alter the resulting morphology? Does it influence the evolution or quantity of tidal material over time?

\begin{figure*}
  \centering
  \includegraphics[width=0.98\linewidth]{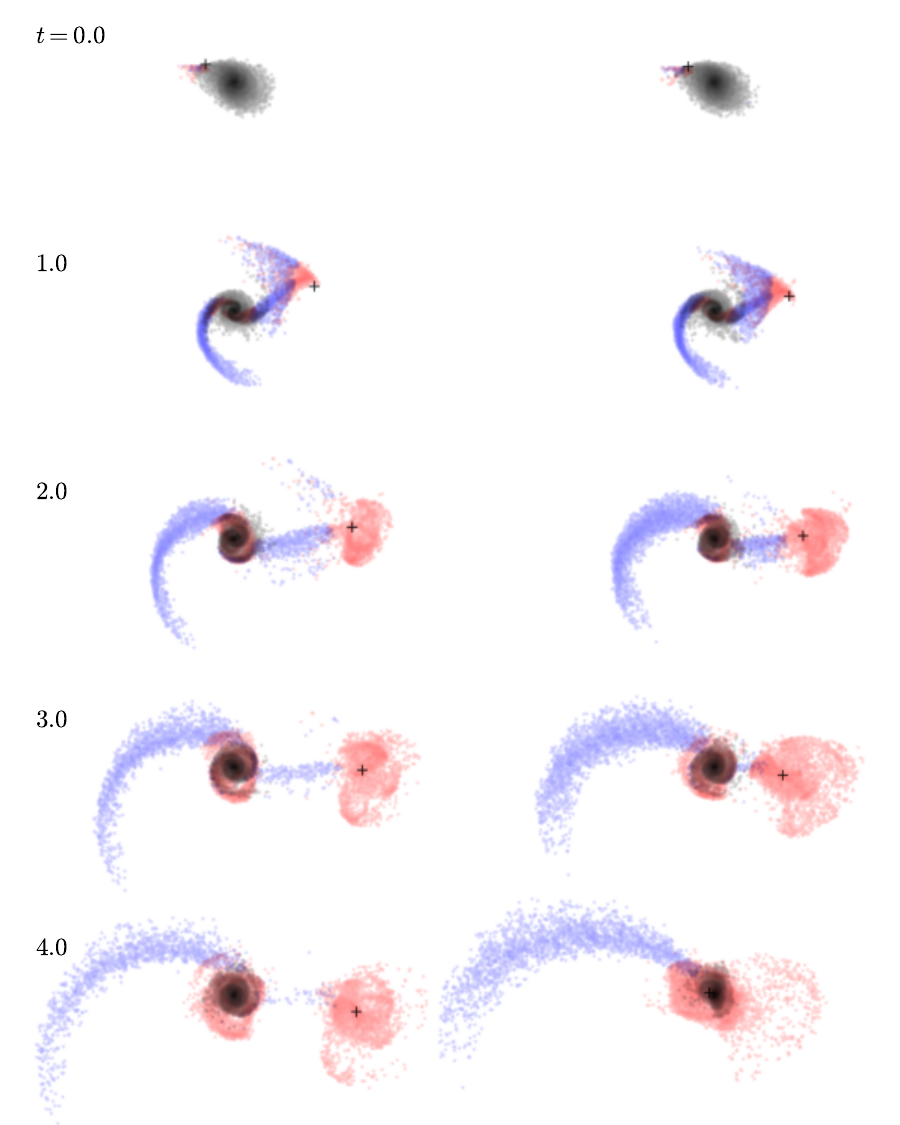}
  \caption{Evolution of the in-plane disc in encounters of isotropic (left) and anisotropic (right) galaxies with pericentric separation $r_\mathrm{p} = 0.5$, and starting time $t_\mathrm{s} = -6$.  Particles in the bridges and tails are shown in blue, while those that have been reaccreted or captured are shown in red.  Crosses indicate the position of the companion galaxy.}
  \label{fig:fig09}
\end{figure*}

Fig.~\ref{fig:fig09} illustrates the key effects that anisotropy has on the evolution of tidal features. Here we show the direct (inclination $i = 0$) disc from encounters of isotropic (left) and anisotropic (right) galaxies with $r_{\mathrm{p}} = 0.5$ and $t_\mathrm{s} = -6$. Particles currently in the bridge and tail are shown in blue, while those which were previously in tidal features but have since been reaccreted or captured are shown in red.

At early times ($t = 1.0$) the discs do not look very different. Both have prominent tidal tails curving to the left, and substantial bridges stretching upward and  to the right. These bridges, which lie in the orbit plane, actually connect the two galaxies; each bridge terminates in a cloud of particles which have been captured by the companion. However, the encounter using anisotropic haloes already exhibits somewhat more massive tidal features.

At later times, the bridge between the anisotropic galaxies  (right) is rapidly consumed as the two galaxies fall back together.  Some material is reaccreted by the parent disc, while the balance is captured by the companion, which is dressed in a growing envelope of tidal debris.  In contrast, the isotropic galaxies  (left) linger near apocentrefor an extended length of time.  During this time, particles at both ends of the bridge are reaccreted or captured; in effect, the bridge is stretched between them.

The evolution of the tails is more straightforward; both continue to grow with time, as Fig.~\ref{fig:fig05} illustrates.  The tails in the encounter of the anisotropic galaxies are broader, longer, and appear more massive than their isotropic counterparts. These differences are likely due to the closer passage of the anisotropic galaxies.

\begin{figure*}
  \centering
  \includegraphics[width=1\linewidth]{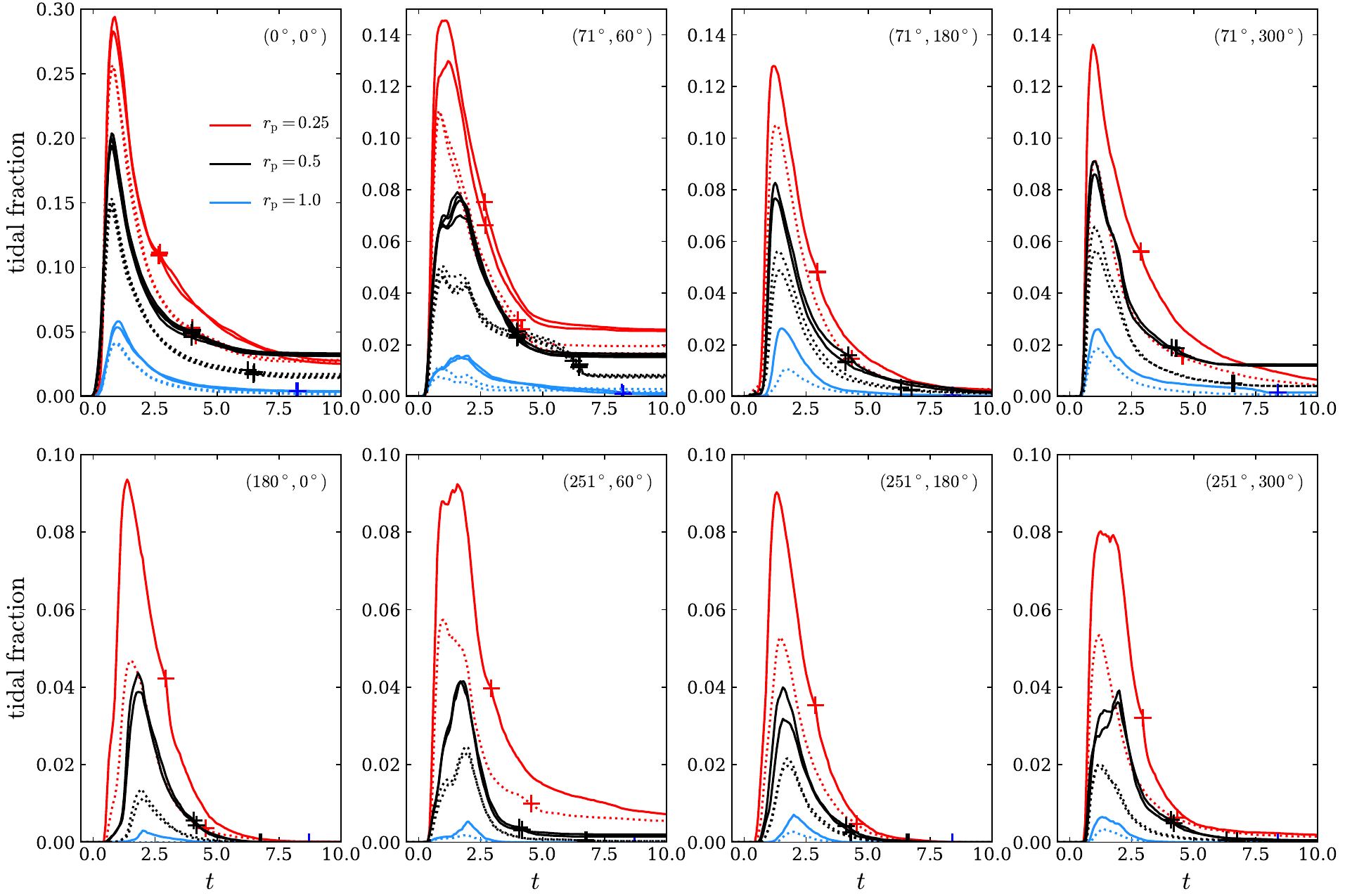}
  \caption{Net tidal fractions as functions of time for encounters of isotropic (dotted) and anisotropic (solid) galaxies. Line color indicates pericentric separation. Each panel shows a different disc orientation as indicated by the angles $(\theta_x, \theta_z)$ in the upper right. The top row depicts direct encounters; note that the first panel has twice the vertical range as the other three.  The bottom row shows retrograde encounters, in effect reversing the spin used in the panel directly above. The markers on the curves show tidal fractions at second pericentre. All simulations starting at $t_\mathrm{s} = -6$ are included.}
  \label{fig:fig10}
\end{figure*}

To show how the amount of tidal material evolves, we plot net tidal fractions versus time in Fig.~\ref{fig:fig10}. The tidal fraction is the number of disc particles  have exceeded their critical radius $r_\mathrm{tid}$ but have not yet fallen back, divided by the total in the disc $N_\mathrm{d}$. Prograde and retrograde encounters are shown on the top and bottom row, respectively; top and bottom pairs spin in opposite directions.

In every case, we observe a rapid increase in the tidal material immediately after first pericentre. This is followed by a steady decline as tidal particles fall back into their parent galaxies, or (in the case of the direct, in-plane discs shown in Fig.~\ref{fig:fig09}) are captured by the other galaxy. Overall, the direct encounters (top row) produce considerably more tidal material than the retrograde ones (bottom row).  Discs with orientation $(0^\circ, 0^\circ)$, which experience the strongest tidal forces as they lie exactly in the orbit plane, yield the most massive tidal features; the other three prograde orientations produce less than half as much. Conversely, all four of the retrograde encounters yield comparable amounts of tidal material, notwithstanding the fact that discs with orientation $(180^\circ, 0^\circ)$ also lie in the orbit plane.

Regardless of disc orientation,  encounters of anisotropic galaxies \textit{always} produce stronger tidal features. The magnitude of this effect depends on circumstances. Anisotropic systems with $r_\mathrm{p} = 0.5$ attain peak tidal fractions between $\sim 30$~and $\sim 200$~per cent higher than their isotropic counterparts, with the in-plane discs (left-hand panels) representing both extremes. For example, the retrograde in-plane discs, $(180^\circ, 0^\circ)$, exhibit the largest increase , while the other three retrograde encounters show slightly smaller fractions. Interestingly, although the prograde disc at $(0^\circ, 0^\circ)$ reaches the highest peak tidal fraction, it has the smallest relative gain  compared to its isotropic equivalent, while the remaining three prograde discs show an average increase of $\sim 50$~per cent. Similar outcomes are seen for the other choices of pericentric separation $r_\mathrm{p}$.

The differences between encounters of isotropic and anisotropic galaxies persist (and become more dramatic) as tidal material falls back.  In Fig.~\ref{fig:fig10}, the markers on the curves indicate tidal fractions at second pericentre.  (No markers are shown for isotropic systems with $r_\mathrm{p} = 1.0$, which don't reach second pericentre until $t > 10$; at such late times, the tidal fractions are effectively zero.)  Prograde encounters of anisotropic galaxies have on average $\sim 270$ per cent more tidal material at second pericentre than their isotropic counterparts; not only do they produce stronger tidal features to begin with, but they also reach second pericentre earlier, so there's less time for tidal material to fall back.

\begin{figure}
  \centering
  \includegraphics[width=.75\linewidth]{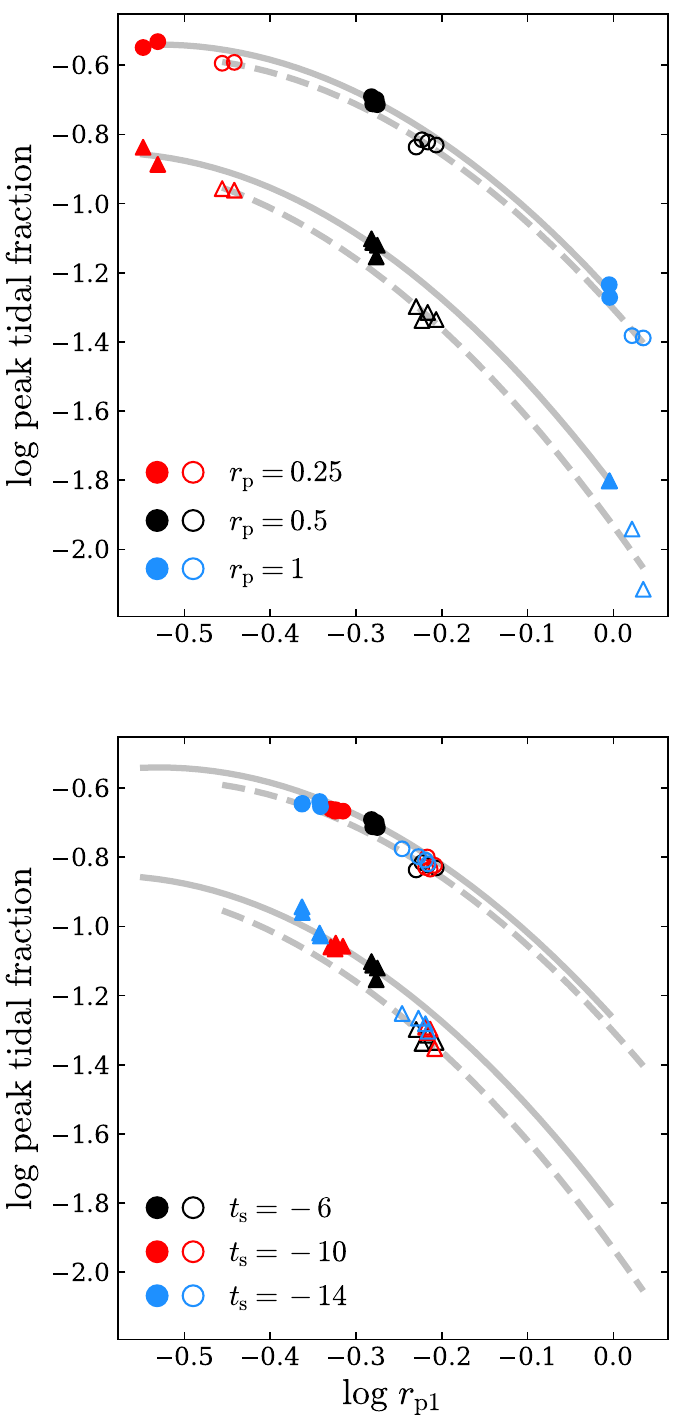}
  \caption{Effects of the idealized pericentric separation (top) and starting time (bottom) on the relationship between the actual pericentric separation $r_\mathrm{p1}$ and the peak tidal fraction. Disc 1 is represented by circles, while disc 2 is represented by triangles. Filled and open symbols distinguish encounters of isotropic and anisotropic systems. All encounters used configuration A; those in the top panel have starting time $t_\mathrm{s} = -6$, while those in the bottom panel have $r_\mathrm{p} = 0.5$. Smooth gray curves are second-order fits to the  data in the top panel.}
  \label{fig:fig11}
\end{figure}

Peak tidal fractions are primarily influenced by actual pericentric separation, as shown in Fig.~\ref{fig:fig11}, while anisotropy has a secondary effect. In the top panel, we plot the peak tidal fractions for the two discs in configuration~A as functions of actual pericentric separation. We observe a consistent trend for each of the disc orientations, with points from the in-plane and inclined discs defining two roughly parallel relationships, here illustrated by second-order fits to the data points. The very close agreement between the relationships for isotropic and anisotropic galaxies with each disc orientation indicates that variations in tidal fraction are primarily driven by differences in pericentric separation.

The bottom panel of Fig.~\ref{fig:fig11} further substantiates this conclusion. Encounters with varying pericentric separations—arising from differences in starting time—result in tidal fractions that follow consistent, orientation-specific trends. These results collectively demonstrate that pericentric separation and  disc orientation are the principal determinants of the peak tidal fraction.

\begin{figure} 
  \centering
  \includegraphics[width=.75\linewidth]{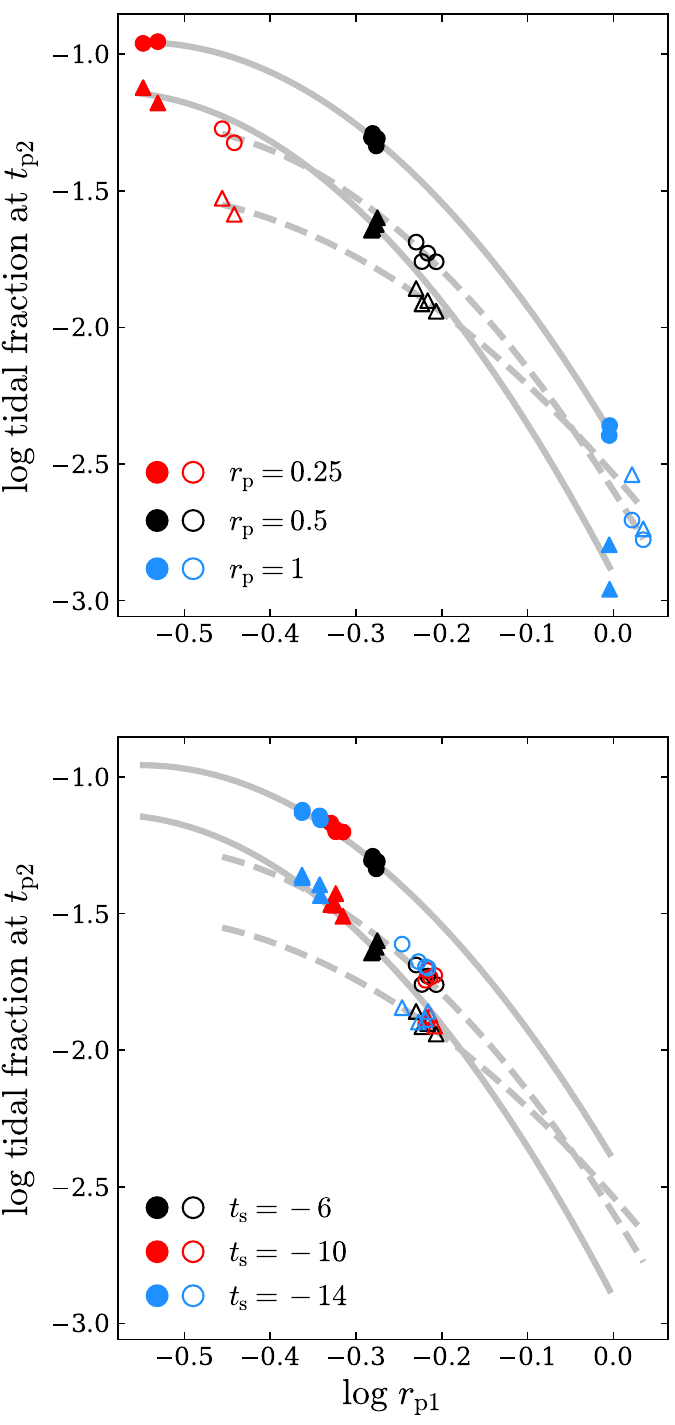}
  \caption{Effects of idealized pericentric separation (top) and starting time (bottom) on the relationship between the actual pericentric separation $r_\mathrm{p1}$ and the tidal fraction remaining at the second pericentre. These are the same encounters illustrated in Fig.~\ref{fig:fig11}.}
  \label{fig:fig12}
\end{figure}

In a similar fashion, Fig.\ref{fig:fig12} illustrates how the tidal fraction at second pericentre (corresponding to the markers in Fig.~\ref{fig:fig10}) varies with pericentric separation. In contrast to the previous figure, the encounters of isotropic (dotted) and anisotropic (solid) galaxies exhibit different relationships. This occurs because anisotropic systems have shorter orbit decay times, so they arrive at second pericentre before much tidal material has fallen back. We note that the results for the widest passages are likely to be somewhat uncertain, as they represent only a handful of particles. 

The bottom panel highlights effects of starting time, which has a small but definite impact on the actual pericentric separation and orbit decay, at least for the anisotropic haloes (\S~\ref{subsec:results_orbit decay}). As a result, late-stage tidal fractions from anisotropic galaxies are larger in simulations which start earlier. In contrast, the tidal material remaining in the isotropic galaxies exhibits no obvious trend with starting time.

\section{Discussion}
\label{sec:discussion}

 Velocity anisotropy influences the dynamics of galaxy encounters  from the very beginning.  Anisotropic  galaxies start interacting earlier during their initial approach, and transfer more angular momentum than their isotropic counterparts. For a given initial orbit, this leads to closer first pericentres and more violent interactions. However, at this stage, the visual differences between the isotropic and anisotropic systems remain subtle (see Fig.~\ref{fig:fig09} at $t= 1.5$); the dynamics that give rise to tidal bridges and tails are similar in the two cases.

In  simulations with anisotropic galaxies, starting earlier gives galaxies more time to transfer angular momentum, leading to even closer first pericentres. In contrast, encounters between isotropic galaxies are much less affected by starting time since they do not tidally deform each other until just before pericentre. This difference is evident in the middle panel of Fig.~\ref{fig:actualPeri_mergeTime}, where the ecounters with anisotropic galaxies are clearly arrayed by starting time, while the isotropic ones show no noticeable pattern. \cite{SB2023} also observed this dependence on starting time. Since tidal forces fall off rapidly with distance, there is presumably some point beyond which starting time no longer matters. However, we can not conclusively say that our earliest starting time of $t_\mathrm{s}=-14$ was early enough to reach that point.

After first pericentre, encounters with closer passages experience shorter orbit decay times, with even shorter times in the case of anisotropic systems. As the left panel of Fig.~\ref{fig:actualPeri_mergeTime} shows, anisotropy continues to influence orbit decay beyond the first passage. If anisotropy had no effect post-pericentre, simulations with isotropic and anisotropic galaxies should exhibit the same relationship between $r_\mathrm{p1}$ and $\Delta t$.

We confirm that disc orientation has only a small effect on orbit decay, with prograde encounters merging slightly faster than their retrograde counterparts.  This result is not very surprising, since $90$~per cent of the mass in our models resides in their haloes. Nonetheless, it's worth noting that the luminous components, despite slightly flattening the anisotropic haloes, do not substantially modify the halo response, which is the main driver of orbit decay.

Isotropic and anisotropic systems both produce bridges and tails, but the way these features evolve is different. At early times, both types of encounters appear morphologically similar.  However, anisotropic  galaxies develop more bridge material -- including the material captured by the other galaxy -- and have more massive tails. These differences arise because the closer interactions between  the anisotropic  galaxies produce stronger tidal forces; in turn, stronger forces extract more prominent tidal features, as previous studies have shown \citep{SW1999, Barnes2016}.  We did not note any morphological features which could unambiguously be attributed to halo anisotropy; while anisotropic haloes are more susceptible to tidal deformation than their isotropic counterparts, the effects of this deformation on visible tidal features are quite subtle.

The amount of tidal material at second passage is presumably a good indication of how much long-lasting material will be left after merger, because merger follows soon after $t_\mathrm{p2}$. Across all our simulations, those with anisotropic galaxies  have higher tidal fractions at second pericentre, as indicated by Fig.~\ref{fig:fig10} and Fig.~\ref{fig:fig12}. This is the result of two factors, larger peak tidal fractions and more rapid orbit decay, leaving less time for tidal material to be reaccreted by the parent galaxy or its companion.  However, the amount of late-stage tidal material also depends on the initial orbit; for example, Fig.~\ref{fig:fig12} shows that systems with anisotropic galaxies  with $r_\mathrm{p} = 0.5$ and isotropic ones  with $r_\mathrm{p} = 0.25$ have comparable tidal fractions at second pericentre.

Our results generalize across different galaxy orientations. In all cases, encounters with anisotropic galaxies--regardless of starting time--lead to closer first passages and shorter orbit decay times, producing more massive tidal features than their isotropic counterparts. But the specific tidal response depends on the galaxies' inclinations, shaping the structure and distribution of these features.
The most massive tidal structures are produced by the in-plane disc in configuration A. In this setup, the disc experiences the strongest tidal forces, leading to more pronounced bridges and tails. In contrast, encounters where the galaxies are inclined experience weaker tidal interactions, resulting in smaller and less prominent tidal features.

\section{Conclusion}
\label{sec:conclusion}

The effects of halo anisotropy in disc galaxy mergers are consistent with the results \cite{SB2023} reported for spherical galaxies: radially anisotropic galaxies are more susceptible to tidal forces, so they begin interacting earlier, transfer orbital momentum more rapidly to internal degrees of freedom, and consequently suffer closer passages and merge faster. These closer passages generate stronger tidal forces between the merging galaxies, leading to the formation of more massive and persistent tidal features.

Why are radially anisotropic systems more responsive than isotropic systems to external tides? This puzzle has been raised several times, without a clear answer.  \cite{RFSMPI2022}, using linear response theory, found that the LMC elicits a stronger response in the MW's stellar halo if the latter is radially anisotropic.  \cite{Vasiliev2024}, using self-consistent N-body simulations, obtained a similar result for the MW's dark halo, and noted that this response enhanced the decay of the LMC's orbit.  Both studies used multi-component galaxy models (like those employed here), complicating the task of understanding the underlying physics.  By way of simplification, \cite{SB2023} adopted spherical models, and showed that radially anisotropic systems also exhibit stronger responses to \textit{impulsive} perturbations.  Since it seems, per \citeauthor{RFSMPI2022}, that self-gravity is not an essential element of this puzzle, the next step may be to analyze spherical test-particle systems subject to tidal impulses.

The degree of anisotropy in the haloes of galaxies has historically been an overlooked parameter. While cosmological simulations indicate that orbital motions are radially anisotropic \citep{WLGM2015}, only a handful of studies have incorporated anisotropic velocity distributions in simulations of galaxy encounters. Our investigation highlights the potential role of anisotropy in these models, even though the actual level of anisotropy is uncertain.

The influence of halo anisotropy on orbit decay adds another complication to the problem of modeling galaxy encounters from observational data \citep{BH2009, Privon2013}.  Absent constraints on halo kinematics, the level of anisotropy must be treated as an unknown.  Moreover, anisotropy and initial orbit are partly degenerate, since radial anisotropy reduces orbital angular momentum during the early stages of an encounter.  If we generally don't know the initial orbits of interacting galaxies, modeling their observed encounters does not provide much leverage on the anisotropy of their haloes.

By increasing the amount of tidal material remaining at late times, halo anisotropy may indeed help to model `twin-tailed' merger remnants such as NGC~7252 \citep{Schweizer1982, HM1995}.  In effect, introducing halo anisotropy loosens constraints on other structural parameters such as disc scale or halo extent \citep{Barnes2016}.  But unless models based on isotropic haloes can be ruled out, which seems unlikely, we can't conclude anisotropic haloes are necessary. The best hope to constrain halo kinematics observationally may come from detailed modeling of Local Group systems. \cite{RFSMPI2022} and \cite{Vasiliev2024} both incorporated halo anisotropy into models of the Milky-Way's interaction with the LMC.  Astrometric missions may eventually provide sufficient information on the distances and proper motions of objects in the haloes of the MW-LMC systems to reconstruct the response of the MW's dark halo to the LMC's tidal field.

Since \cite{TT1972}, simulations of interacting galaxies have become more and more realistic, and including halo anisotropy can be considered a further step along this path.  Yet our results also point to a larger issue with modeling galaxy encounters: the haloes used in our simulations are, \textit{at best}, schematic and incomplete.  Unlike our models, real haloes are almost certainly not in equilibrium at larger radii; galaxies in field and group environments probably accrete material from their surroundings and incorporate it into their haloes via phase mixing over multiple internal orbits \citep{SNRT2020}.  Building a pair of equilibrium objects and staging an encounter between them fails to represent this actual state of affairs.  We can imagine `grafting' infall solutions \citep[e.g.,][]{FG1984, Bertschinger1985} onto our equilibrium models, and then modifying these infall solutions to reflect the presence of two galaxies, but it's not clear that such a project is feasible.  Cosmological simulations which build structure from nearly uniform initial conditions may be the only way to construct plausible halo models for interacting systems; preliminary experiments along similar lines \citep[][Ch.~5]{Blumenthal2019} suggest that much remains to be learned before realistic simulations of observed systems can be produced in this fashion.  Meanwhile, idealized simulations have a role in revealing the physics of galaxy encounters.

\section*{Acknowledgments}
We would like to thank Lucas Blair for helping to develop the algorithms we used for tidal analysis. TMU also thanks the members of the 2024 ASTR 494 Senior Research class at UH M{\=a}noa for their comments and suggestions.  JEB thanks Atsushi Taruya and the Yukawa Institute for hospitality during the revision of this paper.  We thank Frank van~den~Bosch for a constructive and stimulating referee's report, and the Institute for Astronomy for covering the publication costs.

\section*{DATA AVAILABILITY}
Simulation data and software are available by request to the corresponding author at \textsf{barnes@hawaii.edu}.

\bibliography{main}{}
\bibliographystyle{aasjournal}

\appendix

\section{ROI IN COMPOUND MODELS}

Since the radial orbit instability (ROI) has not been extensively studied in compound galaxy models, we present some further details of the results of our stability tests.

The presence of a massive disc should theoretically increase the flattening of the halo. However, if the haloes are stable, we expect (a) that this flattening would reach equilibrium within only a few dynamical times, and (b) that the haloes will remain nearly oblate. This is indeed what we observe for a halo with $r_\mathrm{a} = 0.5$ and a disc comprising $7.5\%$ of the galaxy's mass. In contrast, haloes with $r_\mathrm{a} = 0.354$ become increasingly flattened over many dynamical times and eventually evolve into nearly prolate configurations.

\begin{figure}
    \centering
    \includegraphics[width=\linewidth]{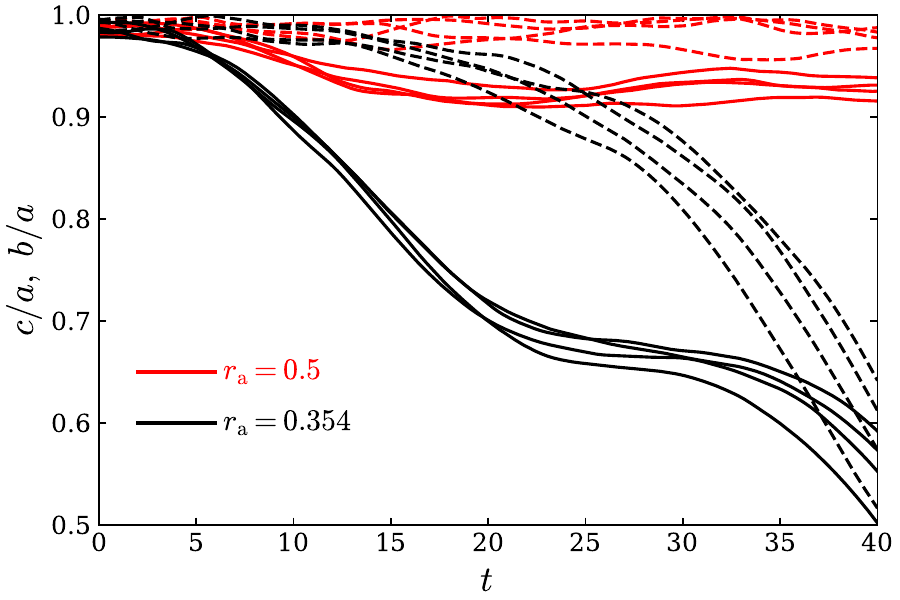}
    \caption{Axial ratios versus time for stable (red) and unstable (black) galaxy models.  Four realizations of each model are shown.  Solid curves are minor-to-major ($c/a$) ratios, while dashed curves are intermediate-to-major ($b/a$) ratios; both are measured for halo particles with binding between the $75^\mathrm{th}$ and $87.5^\mathrm{th}$ percentiles in initial binding energy.}
    \label{fig:figA1}
\end{figure}

\begin{figure*}
  \centering
  \includegraphics[width=0.8\linewidth]{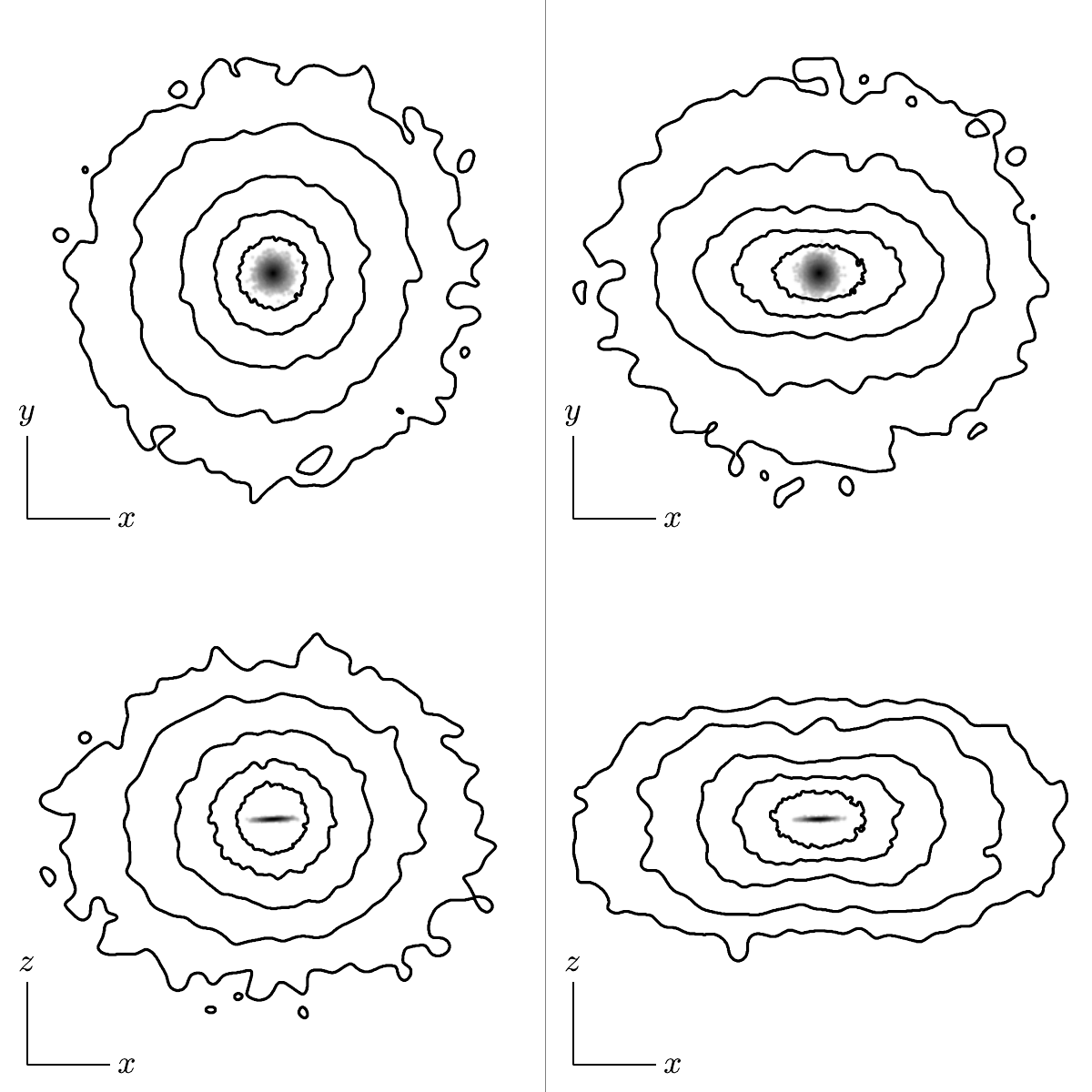}
  \caption{Stable (left) and unstable (right) galaxy models at time $t = 40$, projected along (top) and perpendicular (bottom) to the disc axis.  Discs are shown in grey-scale, while haloes are represented by contours, spaced by a factor of $4$ in projected density; bulges are not shown.  The unstable system has been rotated about the $z$ axis to align the halo's bar with the $x$ axis.  Each frame is $20$ by~$20$ length units.}
  \label{fig:figA2}
\end{figure*}

Fig.~\ref{fig:figA1} displays the time evolution of the axial ratios $c/a$ and $b/a$ for four marginally stable models with $r_\mathrm{a} = 0.5$ (red), and four unstable models with $r_\mathrm{a} = 0.354$ (black). These results extend those shown in Fig.\ref{fig:fig02} by including both axial ratios. The runs with $r_\mathrm{a} = 0.5$, which we believe to be stable, flatten moderately during the first $\sim 15$ time units, and remain nearly oblate, with $b/a \gtrsim 0.95$. In contrast, the runs with $r_\mathrm{a} = 0.354$ gradually evolve toward nearly prolate configurations.

For purely spherical initial conditions, the orientation of an ROI bar presumably depends on small perturbations seeded by $\sqrt{N}$ noise. These are completely random, so the resulting bars have random orientations. However, the presence of a disc breaks spherical symmetry and determines\textemdash at least initially\textemdash the orientation of the ROI. Think of a pencil balanced momentarily on its point: absent any external disturbances, it falls in a random direction, but the slightest breeze is enough to ensure the pencil falls with its point facing upwind. Moreover, the pencil falls somewhat sooner in a breeze than it would in a vacuum.

Fig.~\ref{fig:figA2} presents projections of models with $r_\mathrm{a} = 0.5$ (left) and $r_\mathrm{a} = 0.354$ (right), evolved until time $t = 40$. We interpret the model on the left as stable, while the model on the right is unstable. Models with $r_\mathrm{a} = 0.354$ rapidly flatten toward the disc plane, and only more slowly\textemdash presumably because the resulting oblate configuration is still unstable\textemdash evolve toward prolate shapes. We expect the resulting bars to be parallel to the disc plane but otherwise randomly oriented, which is consistent with what we observe.

Of course, our expectations are based an intuitively plausible analogy, rather than an actual calculation. A proper theory might begin by identifying the growing modes of the ROI in an anisotropic bulge/disc/halo system, and then show how the perturbation provided by the disc couples to these modes, but that is beyond the scope of this paper.

By way of comparison, we also ran stability tests on pure $n = 3.5$ Einasto haloes.  With our methodology, we saw fairly clear indications of the ROI for $r_\mathrm{a} \le 0.4585$.  However, our tests with $r_\mathrm{a} = 0.5$ were ambiguous; two realizations seemed to be slowly but steadily evolving toward more elongated shapes, while the other two may simply have been fluctuating about spherical equilibria.  Simulations with larger $N$, which would be less afflicted by Monte-Carlo uncertainties, might help resolve this ambiguity.  We conjecture that a central bulge and disc, neither of which are radially anisotropic, might act to suppress the ROI in an otherwise marginal case, but our data are not good enough to support such a claim.

\label{lastpage}

\end{document}